\documentclass[lettersize,journal]{IEEEtran}
\usepackage{amsmath}
\usepackage{array}
\usepackage{textcomp}
\usepackage{stfloats}
\usepackage{url}
\usepackage{verbatim}
\usepackage{graphicx}
\usepackage{cite, amsmath, amssymb,amsfonts, algpseudocode, algorithm, algorithmicx, graphicx, textcomp, xcolor, subcaption, thm-restate, amsthm}
\usepackage{graphicx, multirow}

\newtheorem{proposition}{Proposition}[section]

\begin{document}

\title{An Extended Kalman Filter for Systems with Infinite-Dimensional Measurements}

\author{Maxwell M. Varley, Timothy L. Molloy, Girish N. Nair
        % <-this % stops a space
\thanks{The first and third authors were supported by the Australian Government, via grant AUSMURIB000001 associated with ONR MURI grant N00014-19-1-2571. }
\thanks{The second and third authors were supported by the Australian Research Council under the Discovery Project DP250100112.}
\thanks{M. M. Varley and G. N. Nair are with the Department of Electrical and Electronic Engineering, University of Melbourne, Parkville, VIC, 3010, Australia. (emails: max.varley@gmail.com, gnair@unimelb.edu.au)}% <-this % stops a space
\thanks{T. L. Molloy is with the CIICADA Lab, School of Engineering, Australian National University, Canberra, ACT 2601, Australia (e-mail: timothy.molloy@anu.edu.au)}}

% The paper headers
\markboth{Journal of \LaTeX\ Class Files,~Vol.~14, No.~8, August~2021}%
{Shell \MakeLowercase{\textit{et al.}}: A Sample Article Using IEEEtran.cls for IEEE Journals}

% Remember, if you use this you must call \IEEEpubidadjcol in the second
% column for its text to clear the IEEEpubid mark.

\maketitle

\begin{abstract}
This article examines state estimation in discrete-time nonlinear stochastic systems with finite-dimensional states and infinite-dimensional measurements, motivated by real-world applications such as vision-based localization and tracking.
%% GN: commented out and control. 
We develop an extended Kalman filter (EKF) %% GN capable of 
for real-time state estimation, with the measurement noise 
%using infinite-dimensional measurements corrupted by noise 
modeled as an infinite-dimensional random field. 
When applied to vision-based state estimation, the measurement Jacobians required to implement the EKF are shown to correspond to image gradients.
This result provides a novel system-theoretic justification for the use of image gradients as features for vision-based state estimation, contrasting with their (often heuristic) introduction in many computer-vision pipelines.
We demonstrate the practical utility of the EKF on a public real-world dataset involving the localization of an aerial drone using video from a downward-facing monocular camera. 
The EKF is shown to outperform VINS-MONO, an established visual-inertial odometry algorithm, in some cases achieving mean squared error reductions of up to an order of magnitude.
\end{abstract}


\section{Introduction}
\IEEEPARstart{I}{n} this paper we focus on state estimation for systems with finite-dimensional states and infinite-dimensional measurements. 
This focus is motivated by vision-based state estimation, control, and localization problems that arise across robotics~\cite{SLAMSurvey,VINS-MONO,Corke2011} and control~\cite{Silveira2020,Silveira2024,VisualObjectTrackingEKF}.
In such problems, the measurements take the form of images with dimensions determined by the camera's resolution (i.e., number of pixels), while the underlying state of interest (e.g., position and orientation) is typically relatively low-dimensional.
With modern cameras offering increasingly high-resolution images, the emergent challenge in many of these problems is how best to estimate a low-dimensional state with arbitrarily high-dimensional measurements.
Traditional approaches from computer vision and robotics for processing high-dimensional measurements rely on (spatial) feature extraction \cite{Corke2011}.
However, such approaches may fail to exploit the dynamics and uncertainty of the state estimates in determining which features to extract or how to weight them in computing a state estimate.
In this paper, we therefore take a different approach by formulating an extended Kalman filter (EKF) capable of processing an entire (infinite-dimensional) image domain, and assigning dynamic weights (via gains) to every pixel based on its contribution to state estimates.

Although Kalman filters and their nonlinear variants, such as the EKF, have long been used for estimation in robotics and control, the vast majority of vision-based filters are feature-based~\cite{KalmanHistory, KalmanIndustry, VisualObjectTrackingEKF, SLAMSurvey, Corke2011}. 
The use of these filters thus typically involves first reducing images to a sparse set of extracted keypoints before applying standard Kalman filter techniques. 
In contrast, our novel EKF is capable of operating directly on dense image data (i.e., pixel intensities) directly in real-time, avoiding the need to extract features, and preserving the image structure. 
A key insight that allows the formulation of our filter, is the modeling of the measurement noise as an infinite-dimensional random field. This allows us to construct a continuous image-domain measurement model that naturally integrates with the structure of an infinite-dimensional EKF. In doing so, we are able to derive a system-theoretic justification for using image gradients in the filter update step. These image gradients are usually introduced heuristically in computer-vision pipelines (cf.~\cite{Corke2011}), but here emerge from the principles of the filter design itself.

A number of different approaches have been employed in early works to derive the Kalman filter for distributed parameter systems~\cite{Tzaf1, Tzaf2, Falb67, Meditch}, although none of these works examined systems with finite-dimensional states and infinite-dimensional measurements as presented here. A survey contextualizing the methods and results of these early derivations is given in~\cite{CurtainSurvey} and a modern, comprehensive examination of control and estimation of distributed parameter systems is given in~\cite{morris2020controller}. In the case of nonlinear distributed parameter systems, the EKF is generally utilized either by reducing the dimensionality of the underlying system before designing the estimator (the early lumping approach)~\cite{ManziesDragan}, or designing an infinite-dimensional distributed parameter EKF and using some discretization scheme for real-world implementation (the late lumping approach)~\cite{KirstenPaper}.

This article makes the following key contributions, extending the optimal linear filter work in~\cite{VarleyACC, VarleyTAC}.
Firstly, the optimal linear filter originally derived in that prior work is generalized to construct an EKF for systems with finite-dimensional states and infinite-dimensional measurements, and with both nonlinear state dynamics and nonlinear measurement equations. 
We provide a derivation of this EKF and establish and interpret the measurement Jacobians that arise within it, with the latter relating to image gradients in the case of image measurements.
We verify the efficacy of the EKF for vision-based state estimation on a real-world dataset.
Specifically, the filter estimates the position, velocity, acceleration, and yaw of an aerial drone equipped with an Inertial Measurement Unit (IMU) as well as optical cameras providing measurements in the form of grayscale downward-facing images. The estimates are evaluated against the ground truth included in the dataset, and the results are compared with the performance of the well-established monocular visual-inertial odometry algorithm VINS-MONO~\cite{VINS-MONO}, showing that our filter generally achieves superior or comparable state estimation performance.

This article is structured as follows. Section~\ref{sec:prelims} presents the notational definitions used throughout this work. Section~\ref{sec:problemform} will define the system model that we will analyze, as well as the assumptions used throughout. Section~\ref{sec:derivationlinear} will describe a linearization of the previously described system model, in preparation for an application of the optimal linear filter derived in \cite{VarleyTAC} to this linearized system. Section~\ref{sec:Algorithm} will present the filter procedure and two results, Proposition \ref{prop:MeasurementJacobian} and Proposition \ref{prop:GramStructure}, which help to simplify the implementation and reduce the computational complexity of our algorithm. Section~\ref{sec:experimentSetup} will give details pertaining to the dataset, as well as how the filter is implemented on the data within, and our chosen measures of filter performance. Using these performance metrics, Section~\ref{sec:results} demonstrates an empirical verification of the proposed filter and compares the results with those of VINS-MONO.

\section{Preliminaries}\label{sec:prelims}
The space of all real-valued random variables with finite second moments is \(\mathcal{H}\). The space of all real-valued random \(n\)-vectors with finite second moment is \(\mathcal{H}^n\).
Both \(\mathcal{H}\) and \(\mathcal{H}^n\) constitute Hilbert spaces~\cite[Ch. 20]{Fristedt}, with inner products corresponding to expectation, namely,
\begin{align*}
    \langle u, v \rangle_\mathcal{H} &\triangleq E\left[uv\right], &&\forall \ u,v\in \mathcal{H}\\
    \langle u, v \rangle_{\mathcal{H}^n} &\triangleq E\left[u^\top v\right], &&\forall \ u,v\in \mathcal{H}^n.
\end{align*}
The induced norms of these spaces are then 
\begin{align*}
    \|u\|_\mathcal{H}&=E\left[u^2\right], \quad &&\forall \  u\in\mathcal{H}\\
    \|u\|_{\mathcal{H}^n}&=E\left[u^\top u\right], &&\forall \ u\in \mathcal{H}^n.
\end{align*}

We shall utilize the multi-dimensional Fourier transform. The multi-dimensional Fourier transform of any \(f\in L_1(\mathbb{R}^d,\mathbb{R})\) is given by \(\mathcal{F}\{f\}\in L_\infty(\mathbb{R}^d,\mathbb{R})\), which follows the standard definition~\cite[Ch. 1]{stein_fourier}
\begin{align}
    (\mathcal{F}\{f\})(\omega)\triangleq \int_{\mathbb{R}^d}f(i)e^{-2\pi j\omega\cdot i}di,\nonumber
\end{align}
where \(\omega\) is a \(d\)-dimensional frequency vector. The inverse Fourier transform is defined by
\begin{align}
    (\mathcal{F}^{-1}\{f\})(i)\triangleq\int_{\mathbb{R}^d}f(\omega)e^{2\pi ji\cdot \omega}d\omega.\nonumber
\end{align}

Finally, let the set of real $n \times n$ symmetric positive definite matrices be $\mathbb{S}^{n}_{++}$.

\section{Problem Formulation}\label{sec:problemform}
We consider a nonlinear discrete-time system
\begin{align}\label{eq:statedynamics}
    \chi_{k+1} = f_k(\chi_k)+w_k,
\end{align}
for time \(k\in \mathbb{N}\) with state vector \(\chi_k\in \mathbb{R}^n\), process noise \(w_k\in \mathbb{R}^n\), and continuously differentiable (potentially time-varying) state transition functions \(f_k\in C^1(\mathbb{R}^n, \mathbb{R}^n)\). 
The states are partially observed via measurements in the form of an infinite-dimensional field.
Specifically, the measurements \(\zeta_k:\mathbb{R}^d\rightarrow \mathbb{R}^m\), are given by
\begin{align}\label{eq:measurmentfunction}
    \zeta_k(i) = g_k(\chi_k, i) + v_k(i)
\end{align}
for \(k\in \mathbb{N}\) and are defined over the domain \( \mathbb{R}^d\), with measurement locations \(i\in \mathbb{R}^d\),  measurement noise \(v_k:\mathbb{R}^d\rightarrow \mathbb{R}^m\), and with (potentially nonlinear and time-varying) measurement functions \(g_k:\mathbb{R}^n\times \mathbb{R}^d \rightarrow \mathbb{R}^m\). We assume that the measurement functions are bounded and absolutely integrable with respect to the measurement domain \(\mathbb{R}^d\) and continuously differentiable with respect to the state vector \(\chi_k\), so that for all \(i\in \mathbb{R}^d\), \(g_k(\cdot, i)\in C^1(\mathbb{R}^n, \mathbb{R}^m)\).

The additive measurement noise \(v_k\) is assumed to be a wide sense stationary (wss) random field with zero mean and a bounded covariance function such that \(\zeta_k(i)\in \mathcal{H}^m\) for all \(i\in \mathbb{R}^d\). The measurement noise and process noise statistics are such that for all \(j,k \in \mathbb{N}\) and all \(i,i'\in\mathbb{R}^d\)
\begin{align}\label{eq:noisestats}
    E\left[w_k\right] &= 0,\qquad \qquad E\left[w_k w_j^\top\right] = Q\cdot\delta_{j-k}\nonumber\\
    E\left[v_k(i)\right] & = 0,\quad \mkern10mu E\left[v_k(i)v_j(i')^\top\right] = R(i-i')\cdot\delta_{j-k}\nonumber\\
    E\left[v_k(i)w_j^\top\right] &= 0.
\end{align}
Here \(Q\in \mathbb{S}_{++}^{n}\) is a positive-definite matrix, \(R:\mathbb{R}^d\rightarrow \mathbb{R}^{m\times m}\) is bounded and absolutely integrable, and \(\delta_{j-k}\) is the discrete impulse function.

\section{Derivation Methodology}\label{sec:derivationlinear}
We derive an EKF for the system with infinite-dimensional measurements defined by \eqref{eq:statedynamics} and \eqref{eq:measurmentfunction} by first linearizing it, and then applying an optimal linear filter to the linearization (in a manner analogous to the derivation of the EKF for systems with finite-dimensional measurements in~\cite[p. 195]{AndersonandMoore}). This approach is not invalidated by the non-standard assumption that the dimension of the state is finite while the dimension of the measurements is infinite, so long as an appropriate linearization can still be performed. We will now demonstrate such a linearization.

We denote the Jacobians of the state \eqref{eq:statedynamics} and measurement \eqref{eq:measurmentfunction} equations, evaluated at \(\bar{x}_k\), by
\begin{align}
    F_k(\bar{x}_k)&=\left.\frac{\partial f_k(\chi_k)}{\partial \chi_k}\right|_{\chi_k=\bar{x}_k}\in \mathbb{R}^{n\times n},\\ \ G_k(\bar{x}_k, i)&=\left.\frac{\partial g_k(\chi_k,i)}{\partial \chi_k}\right|_{\chi_k=\bar{x}_k}\in \mathbb{R}^{m\times n}\label{eq:G}.
\end{align}

For each time index \(k\), we will denote our filter's \textit{a priori} state estimate by \(\hat{x}_{k|k-1}\), and the (\emph{a posteriori}) state estimate by \(\hat{x}_k\). The state update will be linearized around \(\hat{x}_{k}\), and the measurement equation will be linearized around \(\hat{x}_{k|k-1}\). 

Assuming that the term \((\chi_k-\bar{x}_k)\) is sufficiently small, such that the
higher-order error terms may be neglected, the system \eqref{eq:statedynamics} and \eqref{eq:measurmentfunction} has a corresponding linearized system given by
\begin{align}
    x_{k+1} &= F_k(\hat{x}_{k})x_k+u_k(\hat{x}_k)+w_k\label{eq:linearstate}\\
    z_k(i) &= G_k(\hat{x}_{k|k-1}, i)x_k+y_k(\hat{x}_{k|k-1},i)+v_k(i),\label{eq:linearmeasurement}
\end{align}
where \(u_k:\mathbb{R}^n \rightarrow \mathbb{R}^n\) and \(y_k:\mathbb{R}^n\times \mathbb{R}^d\rightarrow \mathbb{R}^m\) are given by
\begin{align}
    u_k(\hat{x}_{k}) =& f_k(\hat{x}_{k})-F_k(\hat{x}_{k})\hat{x}_{k},\\
    y_k(\hat{x}_{k|k-1},i) =& g_k(\hat{x}_{k|k-1},i)-G_k(\hat{x}_{k|k-1}, i)\hat{x}_{k|k-1}.
\end{align}

Given the linearized state and measurement models \eqref{eq:linearstate}-\eqref{eq:linearmeasurement}, we adopt the standard EKF framework for producing state estimates and their associated matrices, as described in \cite{AndersonandMoore}.

This completes the required setup of a linearized system. In the following section we apply a linear filter, initially derived in~\cite{VarleyACC, VarleyTAC}, to this system and present the resulting EKF.

\section{Proposed Infinite-Dimensional \\Measurement EKF}\label{sec:Algorithm}

The combination of \eqref{eq:linearstate} and \eqref{eq:linearmeasurement} constitutes a linear time-varying system. When corrupted by additive noise with statistical properties given by \eqref{eq:noisestats}, an optimal filter for this system is derived in~\cite{VarleyTAC}. 
Thus, by applying the linear filter presented in~\cite{VarleyTAC} to the linear time-varying system \eqref{eq:linearstate} and \eqref{eq:linearmeasurement}, we arrive at the EKF for systems with infinite-dimensional measurements presented in Algorithm \ref{alg:EKF}.
\begin{algorithm}[h]
\caption{Infinite-Dimensional Measurement EKF}\label{alg:EKF}
\begin{algorithmic}[1]
\State Inputs: \(P_{0} \in \mathbb{S}_{++}^{n}, \hat{x}_{0}, Q \in \mathbb{S}_{++}^n, R(i,i'), f_k(\cdot), g_k(\cdot, i)\)
\For{\(k\geq 1\)}
\State \(\hat{x}_{k|k-1}=f_k(\hat{x}_{k-1})\)
\State \(\phi_k(i)=\mathcal{F}^{-1}\{\bar{G}_k(\hat{x}_{k|k-1},\omega)^\top\bar{R}(\omega)^{-1}\}\)
\State \(S_k = \int_{\mathbb{R}^d} \phi_k(i)G_k(\hat{x}_{k|k-1}, i)\,di\)
\State \(P_{k|k-1}=F_k(\hat{x}_{k-1})P_{k-1}F_k(\hat{x}_{k-1})^\top+Q\)
\State \(P_k= P_{k|k-1}\left(I+S_k P_{k|k-1}\right)^{-1}\)
\State \(\kappa_k(i)=P_k\phi_k(i)\)
\State Obtain measurement \(\zeta_k(i)\)
\State \(\tilde{z}_k(i) = \zeta_k(i)-g_k(\hat{x}_{k|k-1}, i)\)
\State \(\hat{x}_k=\hat{x}_{k|k-1}+\int_{\mathbb{R}^d}\kappa_k(i)\tilde{z}_k(i)\,di\)
\EndFor
\State Outputs: \(\{P_1,P_2,...,P_N\}, \{\hat{x}_1,\hat{x}_2,...,\hat{x}_N\}\)
\end{algorithmic}
\end{algorithm}

 Note that the derivation of the linear filter in \cite{VarleyTAC} includes sufficient conditions that must be imposed on the system for the filter to be rigorously valid in the linear case. Due to the inherent nonlinearity of the system considered here, these conditions may not be satisfied, and optimality guarantees derived for linear systems in \cite{VarleyTAC} may not hold. 
 However, it is possible to verify certain assumptions such that the gain function \(\kappa_k\) and integral given in line 11 of Algorithm \ref{alg:EKF} are well-defined. These assumptions are given in Appendix~\ref{app:systemAssumptions}.

It should be noted that, as may have been expected, Algorithm~\ref{alg:EKF} bears a number of similarities to the finite-dimensional Kalman filter. Indeed, if the infinite-dimensional components such as \(G, R,\) and \(\phi\) are replaced with their finite-dimensional counterparts and matrix multiplication substituted for integration, then the original Kalman filter is recovered with some algebraic manipulation. These algebraic reformulations are necessary primarily due to the fact that we do not propose an equivalent infinite-dimensional inverse operator to the term \((G_k(\hat{x}_{k|k-1}, i)P_kG_k(\hat{x}_{k|k-1}, i')^\top+R(i,i'))\). While such an operator has been considered~\cite{Omatu}, our approach exploits the connection between Toeplitz matrix inversion and the Fourier transform, which mitigates potential computational difficulties involving higher derivatives of the Dirac delta function (see~\cite{VarleyTAC} for a more detailed discussion on this point). 

We now turn our attention to interpreting the structure of the measurement Jacobians \(G_k(\hat{x}_{k|k-1}, i)\) and the innovation covariance matrix \(S_k\) in the infinite-dimensional measurement EKF of Algorithm \ref{alg:EKF} in cases where the measurements are generated by a camera-type mapping and corrupted by ideal white measurement noise.
Let us therefore consider a nonlinear measurement model defined over the measurement domain, as a special case of \eqref{eq:measurmentfunction},
\begin{align}
    \zeta_k(i) = C(p(\chi_k, i)) + v_k(i), \label{eq:measurement_model}
\end{align}
where \( p: \mathbb{R}^n \times \mathbb{R}^d \rightarrow \mathbb{R}^d \) and \( C: \mathbb{R}^d \rightarrow \mathbb{R}^m \) are differentiable mappings.
These mappings describe a grayscale camera viewing a planar scene when $d = 2$ and $m = 1$ with \( p: \mathbb{R}^n \times \mathbb{R}^d \rightarrow \mathbb{R}^d \) mapping image points to world coordinates, and \( C: \mathbb{R}^d \rightarrow \mathbb{R}^m \) mapping world coordinates to grayscale intensity.
Denote the Jacobians associated with these transforms by
\begin{align}
    J_c(p) &= \frac{\partial C(p)}{\partial p} \in \mathbb{R}^{m \times d}, \label{eq:Jc}\\
    J_p(\chi_k, i) &= \frac{\partial p(\chi_k, i)}{\partial \chi_k} \in \mathbb{R}^{d \times n}. \label{eq:Jp}
\end{align}
The following proposition establishes the structure of the measurement Jacobians \(G_k(\hat{x}_{k|k-1}, i)\).

\begin{proposition}[Measurement Jacobian Structure]\label{prop:MeasurementJacobian}
Let the measurement model be given by~\eqref{eq:measurement_model} with differentiable functions \(C\) and \(p\), and let \(\hat{x}_{k|k-1}\) be the predicted state at time \(k\). Then,
\begin{align}
    G_k(\hat{x}_{k|k-1}, i) = J_c(\hat{p}_{k|k-1}) J_p(\hat{x}_{k|k-1}, i), \label{eq:GkJacobian}
\end{align}
where \(\hat{p}_{k|k-1} = p(\hat{x}_{k|k-1}, i)\). 
Furthermore, if \( (m = 1) \), then:
\begin{align}
    G_k(\hat{x}_{k|k-1}, i) = \nabla C(\hat{p}_{k|k-1})^\top J_p(\hat{x}_{k|k-1}, i). \label{eq:GkScalar}
\end{align}
\end{proposition}
\begin{proof}
The chain rule of differentiation gives that
\begin{align*}
    G_k(\hat{x}_{k|k-1}, i) &= \frac{\partial \zeta_k(i)}{\partial \chi_k} \\
           &= \frac{\partial C(p(\chi_k, i))}{\partial \chi_k} \\
           &= \frac{\partial C(p)}{\partial p} \frac{\partial p(\chi_k, i)}{\partial \chi_k} \\
           &= J_c(p(\chi_k, i)) J_p(\chi_k, i).
\end{align*}
Substituting the predicted state \(\hat{x}_{k|k-1}\) and the corresponding coordinates \(\hat{p}_{k|k-1}\) gives the result in~\eqref{eq:GkJacobian}. The scalar case~\eqref{eq:GkScalar} follows immediately from the definition of the Jacobian and the gradient for scalar-valued functions.
\end{proof}

Our second proposition establishes the structure of \(S_k\).

\begin{proposition}[Matrix Structure]\label{prop:GramStructure}
Adopt the hypotheses of Proposition \ref{prop:MeasurementJacobian}.
In addition, assume the measurement noise \(v_k(i)\) is zero-mean, ideal white noise with covariance \(E\left[v_k(i)v_k(i')^\top\right]\) given by
\[
R(i-i') =\Sigma \delta(i-i'),
\]
where \(\Sigma \in \mathbb{R}^{m \times m}\) is constant and positive definite. Then the matrices \(S_k\) are given by
\begin{align}
    S_k = \int_{\mathbb{R}^d} G_k(\hat{x}_{k|k-1}, i)^\top \Sigma^{-1} G_k(\hat{x}_{k|k-1}, i) \, di. \label{eq:SkIntegral}
\end{align}
with the structure of \(G_k(\hat{x}_{k|k-1}, i)\) from~\eqref{eq:GkJacobian} further implying that
\begin{small}
\begin{align}
    S_k = \mkern-3mu\int_{\mathbb{R}^d} \mkern-3muJ_p^\top(\hat{x}_{k|k-1}, i) J_c^\top(\hat{p}_{k|k-1}) \Sigma^{-1} J_c(\hat{p}_{k|k-1}) J_p(\hat{x}_{k|k-1}, i) \, di. \label{eq:SkExpanded}
\end{align}
\end{small}
\end{proposition}
\begin{proof}
   The second part of the proposition follows from the fact that \(\mathcal{F}\{\Sigma \delta(\tau)\}=\Sigma\) and the definition of \(S_k\) given in Procedure~\ref{alg:EKF},
    \begin{align*}
        S_k &= \int_{\mathbb{R}^d}\phi_k(i)G_k(\hat{x}_{k|k-1}, i)\,di\\
        &=\int_{\mathbb{R}^d}\mathcal{F}^{-1}\{\bar{G}_k(\hat{x}_{k|k-1},\omega)^\top \bar{R}^{-1}\}G_k(\hat{x}_{k|k-1}, i)\,di\\
        &= \int_{\mathbb{R}^d}\mathcal{F}^{-1}\{\bar{G}_k(\hat{x}_{k|k-1},\omega)^\top\}\Sigma^{-1}G_k(\hat{x}_{k|k-1}, i)\,di\\
        &= \int_{\mathbb{R}^d}G_k(\hat{x}_{k|k-1}, i)^\top\Sigma^{-1}G_k(\hat{x}_{k|k-1}, i)\,di.
    \end{align*}
    Substitution of the Jacobian structure of \(G_k(\hat{x}_{k|k-1}, i)\) as given in \eqref{eq:GkJacobian} completes the proof.
\end{proof}

We will next make use of Propositions \ref{prop:MeasurementJacobian} and \ref{prop:GramStructure} in Section~\ref{subsec:sysmodel} to apply our EKF to the vision-based state estimation of an aerial drone using images from a downward-facing camera.
However, here we highlight that Proposition \ref{prop:MeasurementJacobian} provides a system-theoretic justification for the use of image gradients in such vision-based state estimation, since when the mappings \( p: \mathbb{R}^n \times \mathbb{R}^d \rightarrow \mathbb{R}^d \) and \( C: \mathbb{R}^d \rightarrow \mathbb{R}^m \) describe a grayscale camera viewing a planar scene (e.g., the pinhole camera model with $d = 2$ and $m = 1$) then $J_c(\cdot) = \nabla C(\cdot)^\top$ is exactly the (spatial) image gradient.
Note also that the key consequence of Proposition \ref{prop:GramStructure} is that ideal white noise completely removes the necessity of computing a Fourier transform at each step.

\section{Experimental Setup}\label{sec:experimentSetup}
In this section, we examine the performance of the EKF in Algorithm~\ref{alg:EKF} for vision-based state estimation. 
Section~\ref{subsec:dataset} will describe the specific problem and a real-world dataset, Section~\ref{subsec:sysmodel} will model the problem for application of our proposed EKF in Section~\ref{subsec:implementation}, and Section~\ref{subsec:performancemetrics} will establish the metrics by which we will evaluate and compare the filter performance. Finally, Section~\ref{sec:results} will present the results and comparative performance of our proposed EKF with VINS-MONO and ground truth.

\subsection{Problem and Dataset}\label{subsec:dataset}
We implement Algorithm~\ref{alg:EKF} to localize an aerial drone using IMU data and grayscale images from a downward-facing camera.
We use data published by the Finnish Geospatial Research Institute (FGI)
Masala Stereo-Visual-Inertial Dataset\cite{FGI-Masala}. This dataset consists of a series of grayscale downward-facing images collected by an aerial drone as it travels along multiple flight trajectories of varying speeds and altitudes. Onboard readings from IMU, as well as ground truth consisting of drone location and orientation over time, are included in the dataset. 

The images were captured at a rate of 15 Hz by two onboard Basler acA2440-75uc cameras, each equipped with a Fujinon HF6XA-5M \(6\)mm lens and facing downward. The original images were taken at a resolution of \(2448\times 2048\) pixels, but were subsampled to a resolution of \(512\times 612\) pixels for real-time processing. The linear acceleration and angular velocity of the drone was measured by an onboard Xsens MTi-680G RTK GNSS/INS, which also provided Real-Time Kinematic (RTK) corrected Global Navigation Satellite System (GNSS) measurements. These readings were recorded at a sample rate of 100 Hz.
See~\cite{MasalaPaper} for additional details about the dataset.

\subsection{System and Measurement Modeling}\label{subsec:sysmodel}
The state of the system at time index \(k\) is represented by an eleven-vector \(x_k=\left[\rho_k, \nu_k, a_k, \theta_k, r_k\right]^\top \in \mathbb{R}^{11}\), where the 3-dimensional position, velocity, and acceleration of the drone are each modeled by the \(3\)-vectors \(\rho_k, \nu_k,\) and \(a_k\) respectively. The yaw and angular velocity of the drone on the horizontal plane are given by the scalars \(\theta_k\) and \(r_k\). In this experiment, we use the East-North-Up (ENU) convention, and the yaw is defined such that 0 radians corresponds to true north and \(\frac{\pi}{2}\) radians corresponds to East. 
The drone system dynamics are represented by the state update equation
\begin{align}\label{eq:stateupdate}
    \begin{bmatrix}
        \rho_{k+1}\\ \nu_{k+1} \\ a_{k+1}\\\theta_{k+1}\\r_{k+1}
    \end{bmatrix}\mkern-3mu=\mkern-3mu\begin{bmatrix}
        I_3 & I_3\Delta t & 0_3 & 0_{3\times 2}\\
        0_3 & I_3 & I_3 \Delta t & 0_{3\times 2} \\
        0_3 & 0_3 & 0_3 & 0_{3\times 2} \\
        0_{2\times 3} & 0_{2\times 3} & 0_{2\times 3} & \mkern-3mu\begin{matrix}
            1 & \Delta t \\ 0 & 0
        \end{matrix}
    \end{bmatrix}\mkern-6mu\begin{bmatrix}
        \rho_{k}\\ \nu_{k} \\ a_{k}\\\theta_{k}\\r_{k}
    \end{bmatrix}\mkern-5mu+\mkern-1muu_k\mkern-1mu+\mkern-1muw_k.
\end{align}
 The vector \(u_k\in \mathbb{R}^{11}\) is the \(3\)-dimensional acceleration and the angular velocity, in the form \(u_k=[0_{1\times 6},u_1,u_2,u_3,0,u_4]^\top\). These are supplied by the accelerometer and gyroscope readings in the IMU frame and are converted to the global (ENU) frame with the aid of estimated orientation data.
 Considerations on modeling the additive process noise \(w_k\in\mathbb{R}^{11}\) are presented in Section~\ref{subsec:implementation}.

The estimator measurements are represented by a pinhole camera model of a 2-dimensional grayscale image, given by the image function 
\(C:\mathbb{R}^2\rightarrow \mathbb{R}\). This function takes two values, describing the planar image location \(i=[i_1,i_2]^\top\), and produces the grayscale intensity \(C(i_1, i_2)\) at that location. 
In accordance with the pinhole camera model of vision, at a position \((\rho^x_k, \rho^y_k)^\top\) and a distance of \(\rho^z_k\) meters from the image, assuming a flat ground plane and true north yaw, this value corresponds to the camera pixel location \(i\cdot \rho^z_kL_f^{-1}\). However, if the ground elevation variation is non-negligible, the distance from the camera must be modified to \(\rho^z_k-e(\rho^x_k,\rho^y_k)\), where \(e:\mathbb{R}^2\rightarrow \mathbb{R}\) is known ground elevation data.

The nonlinear measurement equation \(\zeta:\mathbb{R}^2\times \mathbb{R}^{11}\rightarrow \mathbb{R}\) is then given by
\begin{align}\label{eq:measurementeq}
    \zeta_k(i,\chi_k)=C\left(p_1, p_2\right)+v_k(i),
\end{align}
where \([p_1,p_2]^\top\) at time \(k\) are the transformed coordinates given by
\begin{align}\label{eq:coordtransform}
    \begin{bmatrix}
        p_1 \\ p_2  \end{bmatrix}&= \begin{bmatrix}
            \rho^x_k \\ \rho^y_k
        \end{bmatrix} +R_{\text{rot}}i\cdot(\rho^z_k-e(\rho^x_k,\rho^y_k))L_f^{-1},\\
        \text{where } R_{\text{rot}}(\theta_k)&=\begin{bmatrix}
            \cos{(\theta_k)} & -\sin{(\theta_k)} \\ \sin{(\theta_k)} & \cos{(\theta_k)}
        \end{bmatrix}.
\end{align}

Assuming that the elevation variation is negligible, the gradient of the image function \(C\) with respect to the world coordinates \([p_1, p_2]^\top\), and the Jacobian of the world coordinates with respect to the state vector \(x_k\)  are, respectively, given by
\begin{align}
    \nabla C(\bar{p}) &= \begin{bmatrix}
        \frac{\partial C}{\partial p_1}(\bar{p}_1, \bar{p}_2)\\ \frac{\partial C}{\partial p_2}(\bar{p}_1, \bar{p}_2)
    \end{bmatrix}\\ 
    J_p(\bar{x}_k,i) \mkern-5mu&=\mkern-5mu\begin{bmatrix}
        I_2 \mkern-6mu& R_{\text{rot}}(\bar{\theta}_k)iL_f^{-1}\mkern-6mu & 0_{2\times6} \mkern-6mu& \frac{\partial R_{\text{rot}}}{\partial \theta}(\bar{\theta}_k)i\bar{\rho}^z_kL_f^{-1} \mkern-6mu&\mkern-6mu 0_{2\times 1}
    \end{bmatrix} .
\end{align}

If we denote \(\hat{p}_{k|k-1}\) as the transformed coordinates of \(\hat{x}_{k|k-1}\) according to \eqref{eq:coordtransform}.
We may employ Proposition~\ref{prop:MeasurementJacobian} to find the measurement Jacobian at time \(k\) as

\begin{align}\label{eq:Gkequation}
    G_k(\hat{x}_{k|k-1}, i) = \nabla C(\hat{p}_{k|k-1})^\top J_p(\hat{x}_{k|k-1})\in \mathbb{R}^{1\times 11}.
\end{align}

Note that separating out the Jacobian \(\nabla C\) is particularly useful when a trustworthy representation of the map (a 2-dimensional image in this context) is available, as the gradient function can be pre-calculated and queried at runtime to reduce the computational burden. Additionally, when the measurement noise is well-approximated by ideal white noise as in Proposition~\ref{prop:GramStructure}, this formulation also gives some intuitive sense of which components of the image are prioritized by the filter, as the filter gain can then be expressed as
\begin{align*}
    \kappa_k(i)=P_kJ_p(\hat{x}_{k|k-1},i)^\top (\nabla C(\hat{p}_{k|k-1})) \Sigma^{-1}\in \mathbb{R}^{11\times 1}.
\end{align*}

\subsection{Implementation and Computational Considerations}\label{subsec:implementation}
The dataset obtained for our state estimation testing is openly available online~\cite{FGI-Masala}. Included in this dataset are a series of \texttt{.bag} files for various drone trajectories. Each \texttt{bag} file contains a sequence of grayscale images captured by the left and right cameras during flight accompanied by a corresponding timestamp. The dataset also contains ground truth files for each trajectory, which give the drones local position and orientation, also timestamped. Finally, the dataset contains sensor parameter files for the IMU and cameras. Each camera's intrinsic matrix and radial-tangential distortion coefficients contained within the camera parameter files were used to undistort the images before they were fed to a Matlab implementation of the filter described in Algorithm~\ref{alg:EKF}.

The input vector \(u_k\in \mathbb{R}^{11}\) allows the estimator to access the linear acceleration and angular velocity as recorded by the onboard IMU. At each time step \(k\), all IMU recordings between step \(k-1\) and \(k\) are integrated, and inertial measurements of linear acceleration and angular velocity are converted from the local IMU frame to the global (ENU) frame utilizing the current drone orientation.
The IMU sensor parameters are included in the dataset, which records the noise density and bias instability for both the accelerometer and gyroscope. These values, as well as how they affect the process noise covariance matrix \(Q\), are included in Table~\ref{tab:filter_params}.

Downward-facing images taken at sufficiently high altitudes may be approximated as subsets of a 2-dimensional map, where the altitude, planar position, and yaw influence the scaling, translation, and rotation, respectively. As two cameras were mounted on the drone and the images captured in synchronization, such a map was constructed via reference images taken by one camera (referred to as the reference camera), and filter measurements were provided by the other camera (referred to as the tracking camera). The expected measurement at each time step \(k\) is then the map constructed by the reference camera images, with the appropriate translations, rotations, and scaling dictated by the expected \textit{a priori} state at time \(k\).

The measurement noise is assumed to be well-approximated by ideal white noise, where the covariance structure is modeled as \(E[v_k(i)v_k(i')^\top]=\Sigma \delta(i-i')\in \mathbb{R}\). The value of \(\Sigma \in \mathbb{R}\) expresses the variance of each pixels noise term and directly affects the filter's relative weighting of the process model as opposed to the measurement.

Prior to calculating the measurement innovation, the images undergo a number of pre-processing steps. Both the predicted image and the recorded image are blurred using a Gaussian filter, which serves to smooth out small-scale noise. The level of blurring introduced in the Gaussian filter also affects the relative impact of small-scale and large-scale features present in the image and is achieved via the built-in \texttt{imgaussfilt} Matlab function. The images are then normalized and histogram equalization is performed by the built-in \texttt{histeq} function to enhance contrast. The final pre-processing step is histogram matching to align the intensity distributions of the two images via the built-in \texttt{imhistmatch} function. 

Once the predicted measurement is obtained from the reference camera map and the current \textit{a priori} state estimate, the gradient term \(\nabla C\) is calculated using the built-in \texttt{imgradientxy} function. \(G_k(\hat{x}_{k|k-1}, i)\) is immediately obtained according to \eqref{eq:Gkequation} and the result can then be used to determine the resulting \(P_k\) and the gain \(\kappa_k\) according to Algorithm~\ref{alg:EKF}. The choice of initial \(P_0\) is given in Section~\ref{sec:results}.

\subsection{Performance Metrics}\label{subsec:performancemetrics}

The two performance metrics by which we evaluate the performance of our filter are the Mean Squared Errors (MSE) of the estimated state vector's position and orientation. The MSE of the estimated position is computed as the squared Euclidean distance between the estimate and true positions.
\begin{align*}
    E_\rho = \frac{1}{N}\sum_{k=1}^N \left(\hat{\rho}_k - \rho_k\right)^\top\left(\hat{\rho}_k - \rho_k\right),
\end{align*}
where \(\hat{\rho}_k\) and \(\rho_k\) denote the estimated and true positions at time step \(k\) respectively. The MSE of the estimated yaw is similarly computed by
\begin{align*}
    E_\theta = \frac{1}{N}\sum_{k=1}^N \left(\hat{\theta}_k - \theta_k\right)^2,
\end{align*}
where the angular difference is calculated modulo \(2\pi\) to account for periodicity. Note that unlike the linear Kalman filtering framework, where minimizing the MSE metric leads to an optimal filter, this is generally not the case when applying the EKF to nonlinear systems. Nonetheless, these metrics can be considered a reasonable approximation for systems that are sufficiently well-approximated by a linear system within the discretized temporal increments. 

\section{Experimental Results}\label{sec:results}

\begin{figure}[t]

    \makebox[0.92\linewidth][r]{\includegraphics[width=0.9325\linewidth]{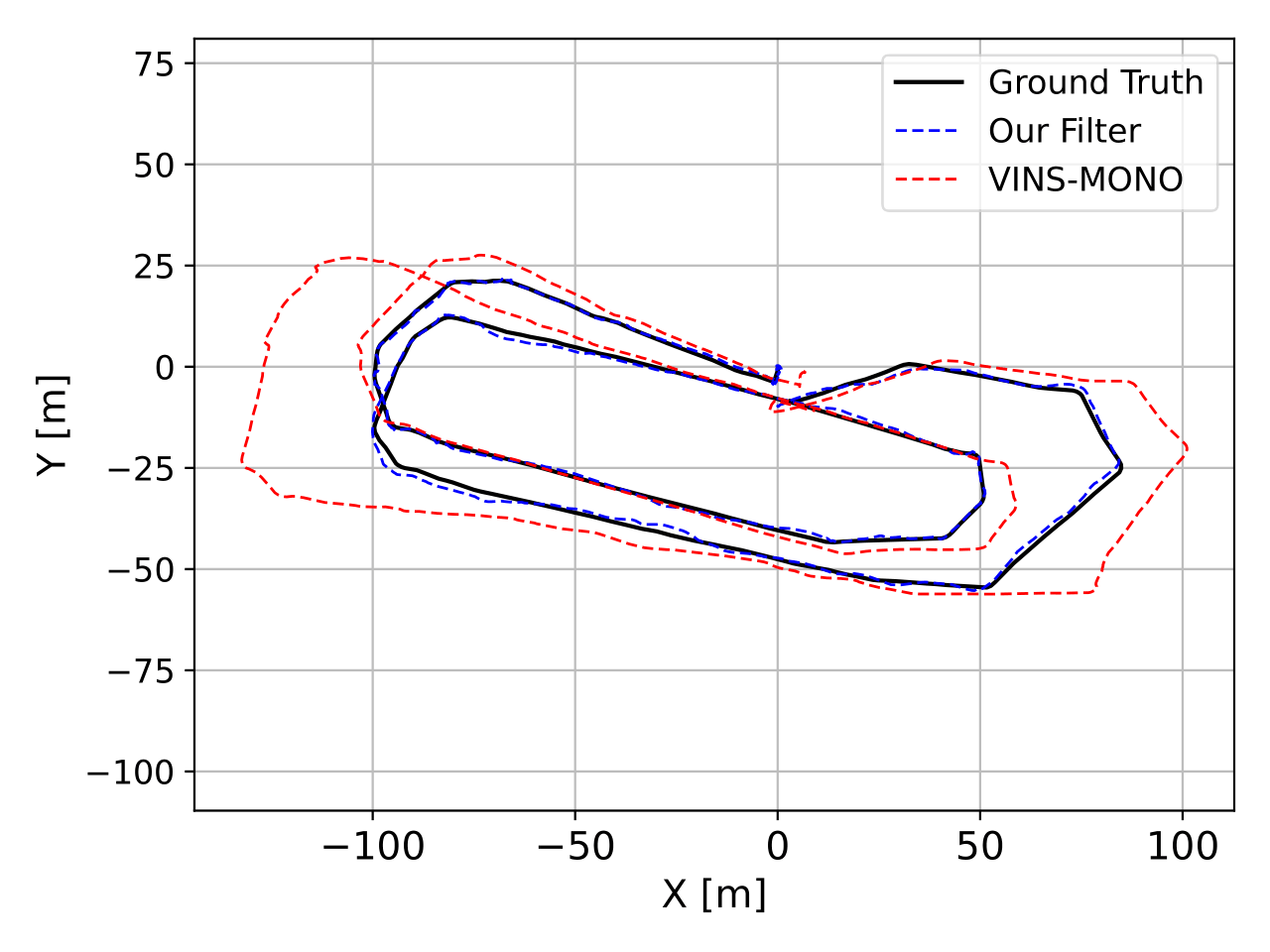}}

    \makebox[0.92\linewidth][r]{\includegraphics[width=0.87\linewidth,height=0.2\textheight,keepaspectratio]{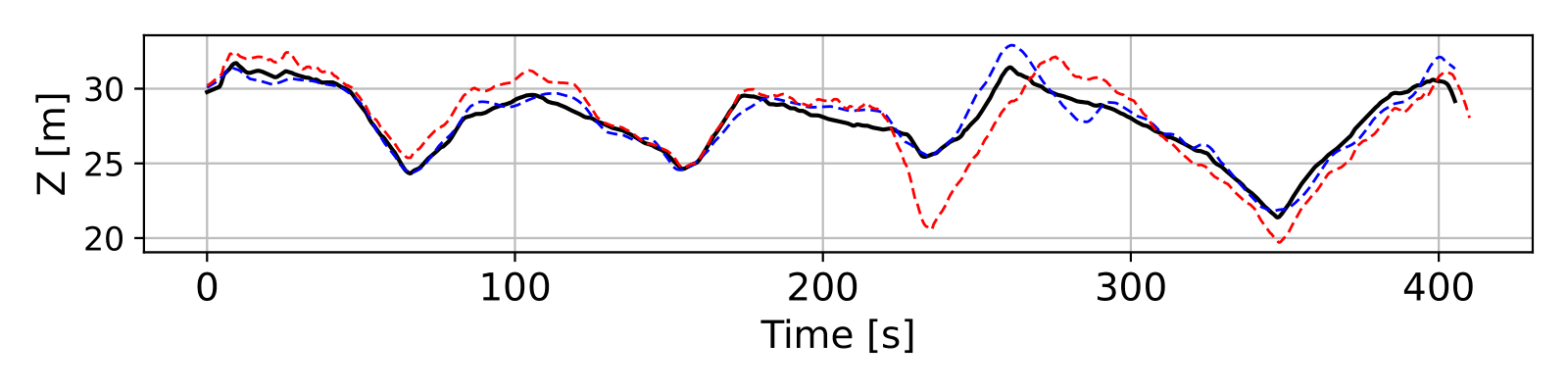}}

    \makebox[0.92\linewidth][r]{\includegraphics[width=0.87\linewidth,height=0.2\textheight,keepaspectratio]{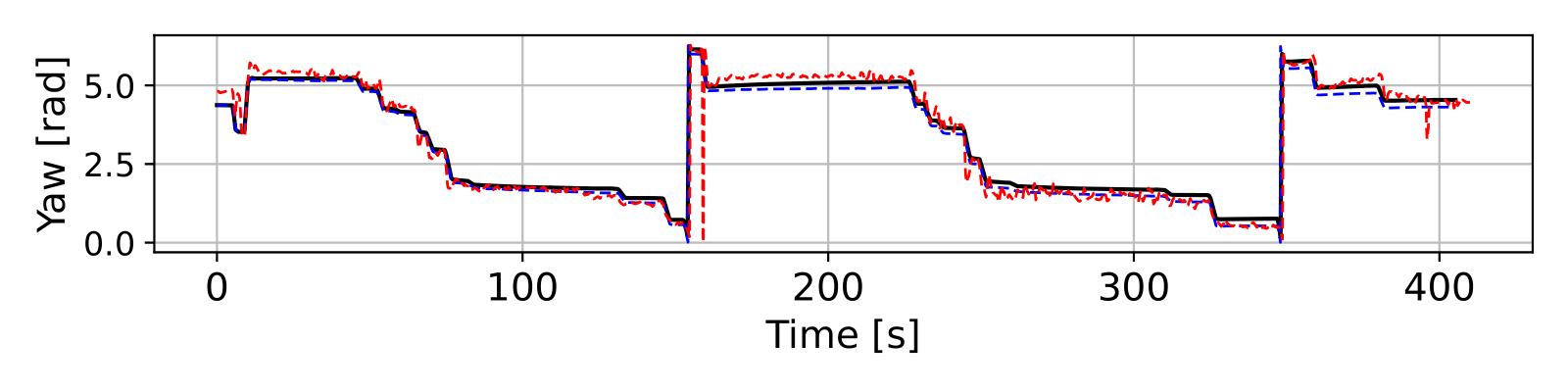}}

    \caption{Planar trajectory (top), altitude over time (middle), and yaw over time (bottom) for the VINS-MONO estimate, our filter estimate, and ground truth. These results are for trajectory \texttt{40\_2} and accelerometer noise density \(1.6 \times 10^{-2}\) m/s\(^2\)/\(\sqrt{\text{Hz}}\).}

    \label{fig:stacked_trajectory}
\end{figure}

Having established the system model and implementation details, we now present the experimental results obtained using our proposed filter on the FGI Masala dataset. To benchmark performance, we compare our filter against VINS-MONO, a feature-based visual-inertial odometry algorithm designed for systems with monocular image measurements. As VINS-MONO has no interface to directly provide ground truth, the algorithm is run over each trajectory in the dataset once using the reference camera images. The resulting map constructed from this reference flight is saved and provided to the subsequent evaluated flight. All comparisons are made against the ground truth position and orientation included in the dataset. All filter and sensor parameters are summarized in Table~\ref{tab:filter_params}, and all are kept constant across each trajectory with the sole exception of the accelerometer noise density parameter. Estimates of covariance matrices for Inertial Navigation Systems (INS) can be found in~\cite[Ch. 12]{Bonnor} and our construction of the process noise covariance matrix \(Q\) is in line with the framework presented there.

Figure~\ref{fig:stacked_trajectory} provides a representative example of the filter performance for the trajectory \(\texttt{40\_2}\). For each trajectory label, the first number denotes the planned altitude of the drone flight in meters, while the number following the underscore indicates the planned average horizontal speed of the flight in meters per second. Across all trajectories in the dataset, the average flight times range from approximately \(6\) to \(10\) minutes, covering a distance of roughly \(800\) to \(1000\) meters. To focus solely on steady-state estimation performance, we omit the initial takeoff and landing segments of each flight. It is also worth noting that the trajectories with different speeds were recorded under varying weather conditions and camera exposure settings. 

In this particular example of trajectory \(\texttt{40\_2}\), a low-altitude and low-speed flight, VINS-MONO and our filter was run with an accelerometer noise density of \(1.6\times 10^{-2}\) m/s\(^2\)/\(\sqrt{\text{Hz}}\). The results show that while neither algorithm diverges, our approach outperforms VINS-MONO in planar estimation, moderately improves performance in altitude estimation, and achieves roughly equivalent performance in yaw estimation. These figures serve as an illustrative case study, but a more comprehensive aggregate analysis across multiple trajectories and noise parameters is presented later in this section.

\begin{table}[t]
    \centering
    \caption{Filter and Sensor Parameters}
    \label{tab:filter_params}
    \begin{tabular}{l|l}
        \hline
        \textbf{Filter Parameters} & \\
        \hline
        Process Noise Covariance \(Q\) & \(I_{11}\mkern-1mu[0_{3\times 1}\mkern-1mu,\mkern-1mu \sigma_a^2\mkern-1mu,\mkern-1mu\sigma_a^2,\mkern-1mu\sigma_a^2\mkern-1mu,\mkern-1mu0,\mkern-1mug_a^2\mkern-1mu,\mkern-1mu0]^\top \mkern-1mu\Delta_t\) \\
        Measurement Noise Covariance \(R\) & \(\Sigma\cdot\delta(i-i')\) \\
        Measurement Noise Magnitude \(\Sigma\) & \(1 \times 10^{-2}\) \\
        Initial Covariance \(P_0\) & \(Q\) \\
        Image Dimensions & \(612\times512\) pixels \\
        Blur Factor & \(0.5\) \\
        \hline
        \textbf{Sensor Parameters} & \\
        \hline
        IMU Frequency & \(100\) Hz \\
        Camera Capture Frequency & \(15\) Hz \\
        Time Step \(\Delta_t\)& \(\frac1{15}\) s \\
        Gyroscope Noise Density \(g_a\)& \(1.94 \times 10^{-3}\) rad/s/\(\sqrt{\text{Hz}}\) \\
        Gyroscope Random Walk \(g_b\)& \(3.96 \times 10^{-5}\) rad/s\(^2\)/\(\sqrt{\text{Hz}}\) \\
        Accelerometer Noise Density \(\sigma_a\) & Variable m/s\(^2\)/\(\sqrt{\text{Hz}}\)\\
        Accelerometer Random Walk \(\sigma_b\)& \(1.31 \times 10^{-4}\) m/s\(^3\)/\(\sqrt{\text{Hz}}\) \\
        \hline
    \end{tabular}
\end{table}

For each trajectory, the variable parameter is the accelerometer noise density, which determines the level of uncertainty in the linear acceleration measurements of the IMU. A lower noise density suggests that the accelerometer readings are relatively trustworthy, whereas higher values suggest a greater presence of acceleration noise. This parameter directly influences the process noise covariance matrix \(Q\), as shown in Table~\ref{tab:filter_params}. The diagonal elements of \(Q\) associated with the velocity estimates are scaled proportionally to \(\sigma_a^2\). Increasing the noise density results in a larger process noise assumption, while decreasing the noise density results in a smaller process noise assumption. The implications for our filter are the same as in the classical Kalman filter framework, a larger \(Q\) causes the filter to rely more heavily on the measurement model, while a smaller \(Q\) increases trust in the process model. Both algorithms performed state estimation on each trajectory \(14\) times. Each experiment exponentially increased the accelerometer noise density parameter, beginning with an initial \(\sigma_a=1\times 10^{-3}\) m/s\(^2\)/\(\sqrt{\text{Hz}}\) and doubling until reaching \(\sigma_a=8.192\) m/s\(^2\)/\(\sqrt{\text{Hz}}\). This method of exponential increase enables efficient exploration of a wide range of accelerometer noise density values without requiring an infeasibly large number of experiments.

Both algorithms were run in real-time, and images were fed to each algorithm at a rate of \(15\)Hz mirroring the capture rate of the drone-mounted camera. The implementation presented in Algorithm~\ref{alg:EKF} involves an inverse Fourier transform at each time step. Although these may be handled computationally by way of the Fast Fourier Transform, this potential bottleneck is avoided under the assumption of zero-mean ideal white measurement noise. In this case, as shown by Proposition~\ref{prop:GramStructure}, the Fourier transform is not necessary and the Fourier calculation is replaced by matrix multiplication according to \(S_k=\int_{\mathbb{R}^d}G_k(\hat{x}_{k|k-1}, i)^\top \Sigma^{-1}G_k(\hat{x}_{k|k-1}, i)\,di\).
\begin{figure}[t]
    \centering

    \begin{subfigure}[t]{0.95\linewidth}
        \centering
        \includegraphics[width=\linewidth]{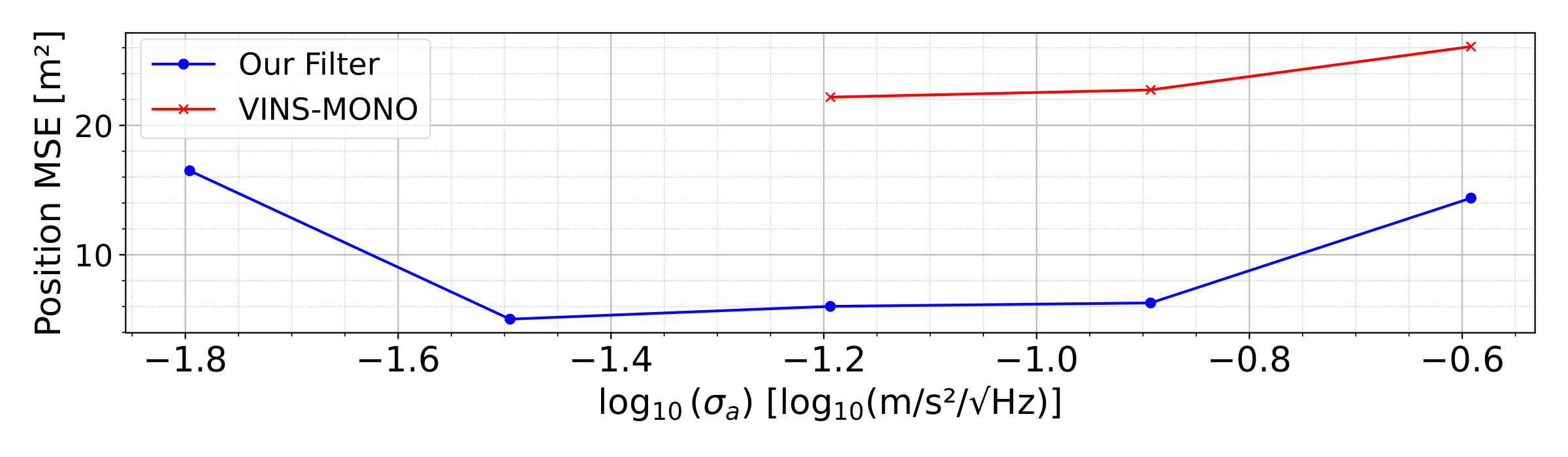}
        \label{fig:avg_mse_2}
    \end{subfigure}

    \begin{subfigure}[t]{0.95\linewidth}
        \centering
        \includegraphics[width=\linewidth]{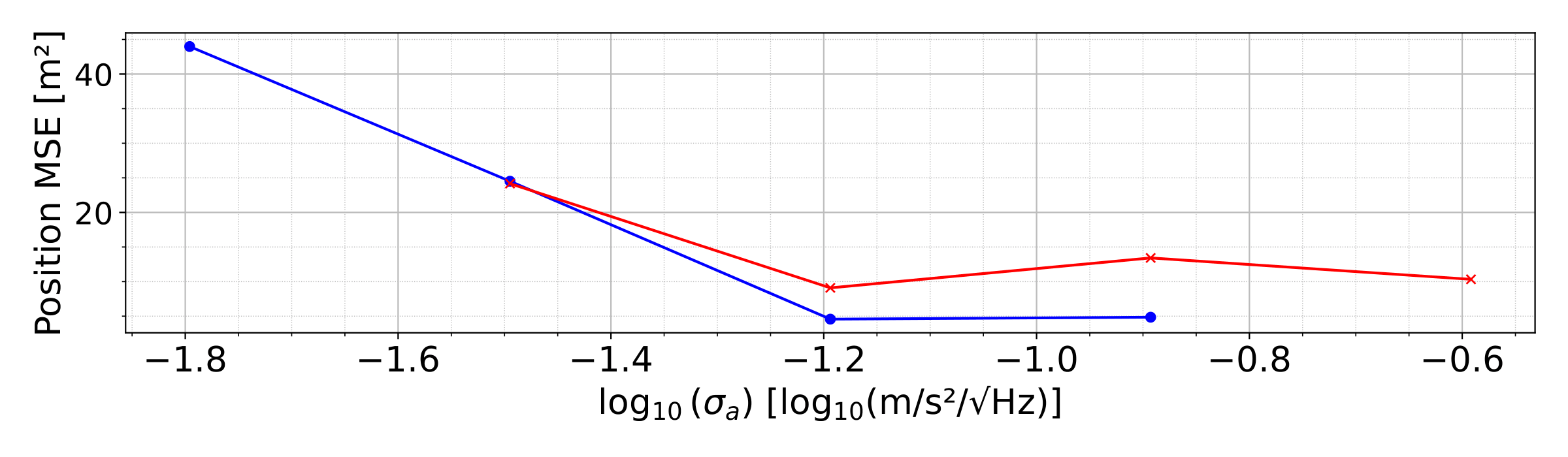}
        \label{fig:avg_mse_3}
    \end{subfigure}

    \begin{subfigure}[t]{0.95\linewidth}
        \centering
        \includegraphics[width=\linewidth]{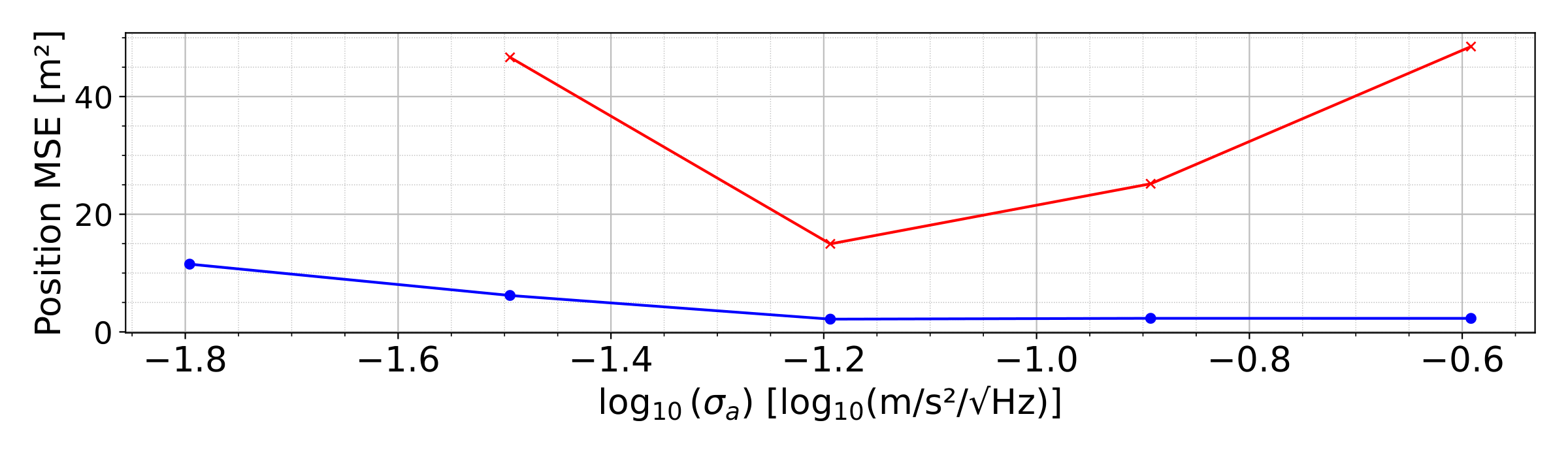}
        \label{fig:avg_mse_4}
    \end{subfigure}

    \caption{Average position MSE across all trajectories, grouped by speed, as a function of accelerometer noise density. Results are shown for speed groups \texttt{2} (top), \texttt{3} (middle), and \texttt{4} (bottom), with values above 100 omitted.}
    \label{fig:avg_mse_stacked}
\end{figure}

For a broader comparison of filter performance, Figure~\ref{fig:avg_mse_stacked} presents the average value of the position MSE on different accelerometer noise densities \(\sigma_a\), grouped by trajectory speed. In the low-speed \(\texttt{2}\) and high-speed \(\texttt{4}\) groups, our filter consistently outperforms VINS-MONO across a wide range of noise accelerometer noise settings. Our filter also shows superior or comparable performance in the medium-speed \(\texttt{3}\) group. However, two trajectories, \(\texttt{40\_3}\) and \(\texttt{100\_3}\), lead to estimator divergence across all tested accelerometer noise levels, with average errors exceeding the threshold of \(100\). These trajectories were excluded from Figure~\ref{fig:avg_mse_stacked}, but included in Table~\ref{tab:best_components}.

\begin{figure}[t]
    \centering

    \begin{subfigure}[t]{0.8\linewidth}
        \centering
        \includegraphics[width=\linewidth]{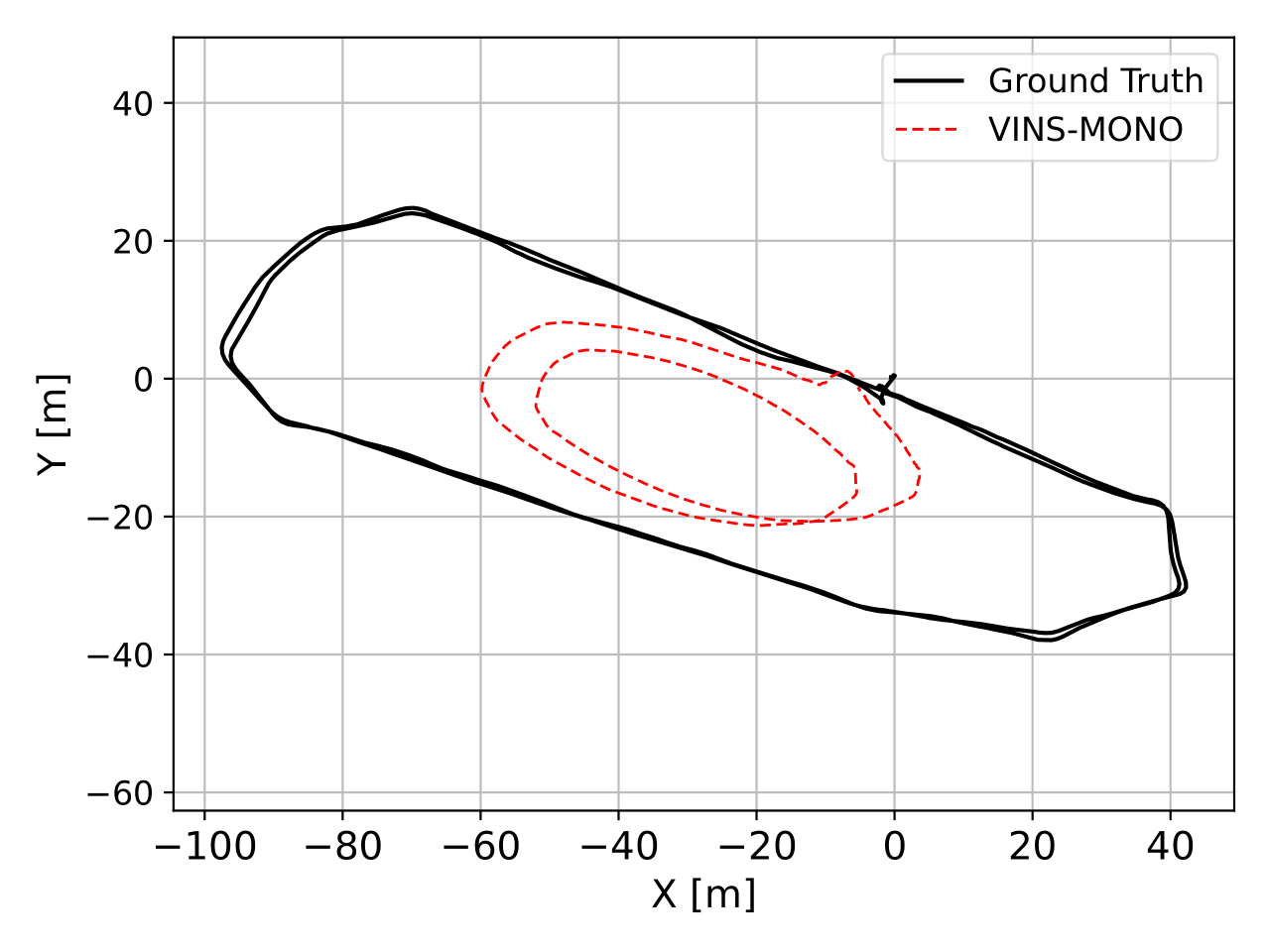}
        \label{fig:vinsfail}
    \end{subfigure}

    \begin{subfigure}[t]{0.8\linewidth}
        \centering
        \includegraphics[width=\linewidth]{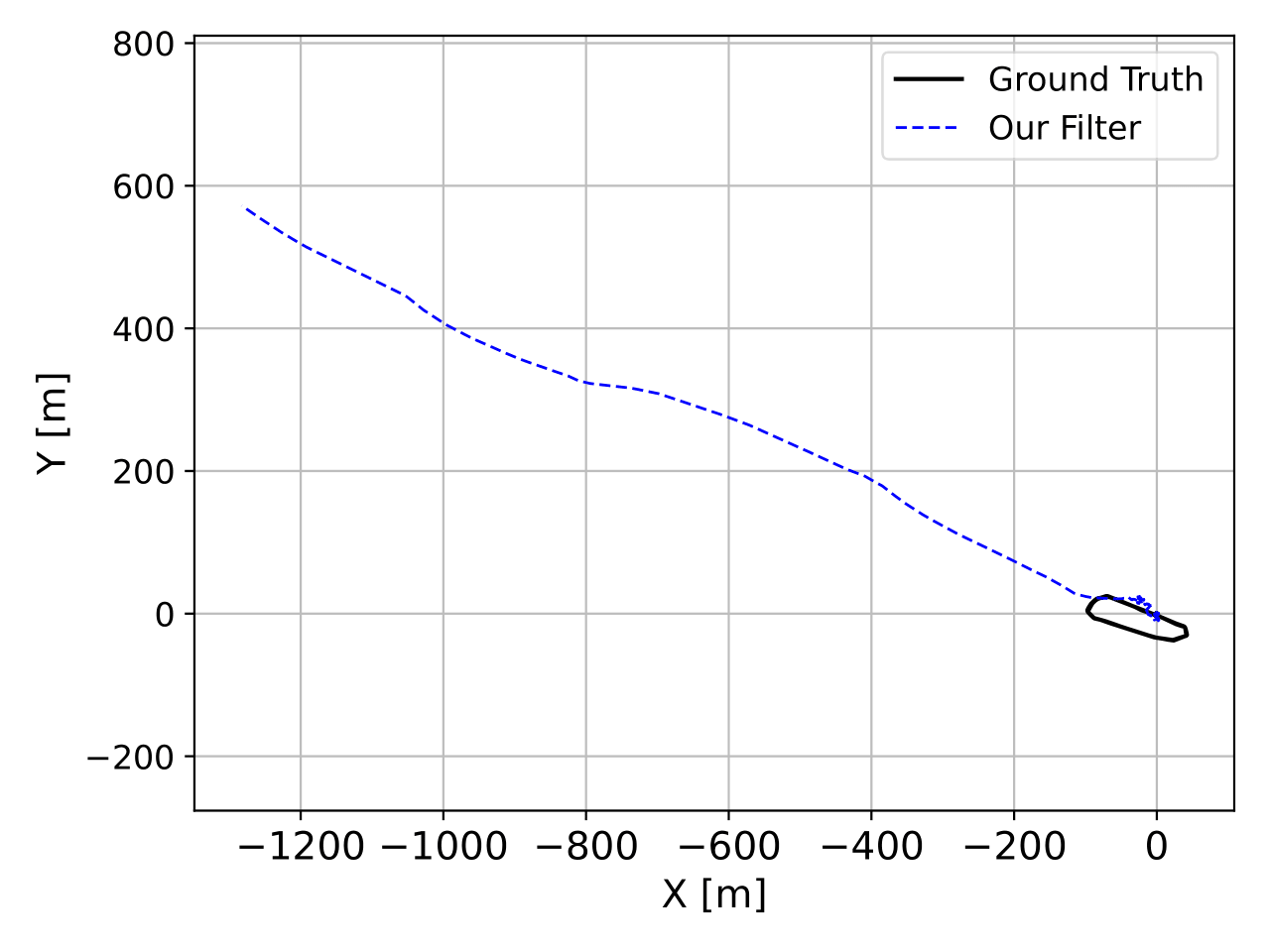}
        \label{fig:ourfail}
    \end{subfigure}

    \caption{Typical failure modes for VINS-MONO (top) and our filter (bottom). Trajectory \(\texttt{100\_3}\) and accelerometer noise density 1.024 m/s\(^2\)/\(\sqrt{\text{Hz}}\).}
    \label{fig:fails}
\end{figure}

Table~\ref{tab:best_components} shows the best MSE values for each filter across all tested accelerometer noise densities. Again, our filter outperforms VINS-MONO on every trajectory except for \(\texttt{40\_3}\) and \(\texttt{100\_3}\), where it experienced divergence before the end of the trajectory for all tested noise values. VINS-MONO remained bounded in these two cases. Yaw accuracy is broadly comparable, across ten non-diverging trajectories our filter attains a lower yaw MSE on four, with VINS-MONO performing better on the remaining six. Typical failure modes are illustrated in Figure~\ref{fig:fails}. When our filter fails, it tends to diverge from the true path with increasing velocity, whereas VINS-MONO tends to spiral into the interior of the trajectory, maintaining bounded position error. 

\begin{table}[t]
\centering
\caption{Position and yaw MSE's for each trajectories best-case accelerometer noise density, lowest MSE in \textbf{bold}.}
\label{tab:best_components}
\begin{tabular}{|l|r|r|r|r|}
\hline
\multirow{2}{*}{\textbf{Trajectory}} &
\multicolumn{2}{c|}{\textbf{Position MSE} [\(m^2\)]}&
\multicolumn{2}{c|}{\textbf{Yaw MSE} [\(rads^2\)]}\\
\cline{2-5}
& \textbf{Ours} & \textbf{VINS-MONO} & \textbf{Ours} & \textbf{VINS-MONO} \\
\hline\hline
40\_2  & \textbf{1.983} & 11.911 & \textbf{0.035} & 0.043 \\
60\_2  & \textbf{2.366} & 5.066 & \textbf{0.014} & 0.032 \\
80\_2  & \textbf{5.197} & 10.410  & 0.020 & \textbf{0.010} \\
100\_2 & \textbf{8.985} & 30.719 & 0.018 & \textbf{0.003} \\
\hline
\hline
40\_3*  & 2013.887 & \textbf{2.614} & 0.033 & \textbf{0.013} \\
60\_3  & \textbf{1.601} & 3.466 & 0.025 & \textbf{0.002} \\
80\_3  & \textbf{7.543} & 11.816 & \textbf{0.034} & 0.049 \\
100\_3* & 60.638 & \textbf{10.223} & 0.064 & \textbf{0.004} \\
\hline
\hline
40\_4  & \textbf{1.357} & 1.402  & \textbf{0.018} & 0.221 \\
60\_4  & \textbf{0.736} & 8.031 & 0.010 & \textbf{0.003} \\
80\_4  & \textbf{1.752} & 17.724 & 0.010 & \textbf{0.008} \\
100\_4 & \textbf{4.610} & 32.464 & 0.011 & \textbf{0.007} \\
\hline
\end{tabular}
\end{table}

\section{Conclusions}

This paper has presented an extended Kalman filter (EKF) for systems with finite-dimensional states and infinite-dimensional measurements, motivated by the challenge of vision-based state estimation. By extending a previously derived optimal linear filter~\cite{VarleyTAC}, we construct a novel nonlinear filtering algorithm via a linearization approach. The effectiveness of this approach was demonstrated through empirical evaluation over a real-world aerial drone dataset. The proposed algorithm is compared with the VINS-MONO estimator, a well-established visual-inertial odometry algorithm. Results across multiple trajectories and noise parameters show that our filter generally achieves superior or comparable estimation accuracy.

Despite this general improvement in estimation accuracy, the proposed filter suffers from degraded performance under poor noise tuning to a greater extent than VINS-MONO. There were also two trajectories in the aerial dataset under which the filter diverged before the end of the trajectory for all tested noise parameters.

There are several promising avenues for future work. Reliance on an existing map limits the current applicability of the proposed filter to unmapped environments. Incorporating online map construction would expand the method from a localization algorithm to a full SLAM algorithm. Also, the process model proposed here assumes planar motion with translation, rotation, and scaling of the image plane, but does not account for skew effects which would arise from non-negligible drone pitch and roll values. Extending the model to capture full SE(3) motion would enable the filter to handle arbitrary rigid-body transformations in three dimensions. Finally, the present formulation does not incorporate IMU bias, which is unique to the individual IMU unit and is typically estimated online by many visual-inertial odometry systems. This can lead to drift over longer time horizons, and addressing this would enhance the robustness of the filter over longer drone trajectories.

\appendices
\section{System Assumptions}\label{app:systemAssumptions}
The following assumptions are sufficient to ensure that the filter derived in~\cite{VarleyTAC} may be applied to the linearized system presented in Section~\ref{sec:derivationlinear}.

\begin{restatable}{assumption}{assumgamma}\label{ass:gamma}
    \(G_k\in L_1(\mathbb{R}^d,\mathbb{R}^{n\times m})\cap L_\infty(\mathbb{R}^d,\mathbb{R}^{n\times m}) \ \forall k.\)
\end{restatable}

\begin{restatable}{assumption}{assumR}
    \label{ass:R} \(R\in L_1(\mathbb{R}^d,\mathbb{R}^{m\times m})\cap L_\infty(\mathbb{R}^d,\mathbb{R}^{m\times m})\).
\end{restatable}

\begin{restatable}{assumption}{assumkappa}
    \label{ass:kappa} \(\kappa_k\in L_1(\mathbb{R}^d,\mathbb{\mathbb{R}}^{n\times m})\cap L_2(\mathbb{R}^d,\mathbb{\mathbb{R}}^{n\times m}) \ \forall k\).
\end{restatable}

\begin{restatable}{assumption}{assumbarR}
    \label{ass:barR} \(\bar{R}(\omega)\in \mathbb{R}^{m\times m}\) is invertible for almost all \(\omega\).
\end{restatable}

\begin{restatable}{assumption}{assumRGamma}
    \label{ass:RGamma} \(\bar{G}_k^\top\bar{R}^{-1}\in L_1(\mathbb{R}^d, \mathbb{R}^{n\times m})\cap L_2(\mathbb{R}^d, \mathbb{R}^{n\times m})\) and admits an inverse Fourier transform in \(L_1(\mathbb{R}^d,\mathbb{R}^{n\times m}) \ \forall k\).
\end{restatable}

\bibliographystyle{ieeetr}
\bibliography{references}

\end{document}